\documentclass[preprint2]{aastex6}
\pdfoutput=1 %for arXiv submission
\usepackage{hyperref}
\usepackage[figure,figure*]{hypcap}

\begin{document}
\title{ Gravitational wave observations may constrain gamma-ray burst models:\\
the case of GW 150914 - GBM}
\author{P. Veres\altaffilmark{1},
R. D. Preece\altaffilmark{2}, 
A. Goldstein\altaffilmark{3,4}, 
P. M\'esz\'aros\altaffilmark{5}, 
E. Burns\altaffilmark{6} and
V. Connaughton\altaffilmark{3}}
\email[e-mail: ]{peter.veres@uah.edu}
\altaffiltext{1}{ CSPAR, University of Alabama in Huntsville, 320 Sparkman Dr.,
Huntsville, AL 35805, USA}
\altaffiltext{2}{Dept. of Space Science, University of Alabama in Huntsville, 320
Sparkman Dr., Huntsville, AL 35805, USA }
\altaffiltext{3}{Universities Space Research Association, 320 Sparkman Dr. Huntsville, AL
35806, USA }
\altaffiltext{4}{Astrophysics Office, ZP12, NASA/Marshall Space Flight Center,
Huntsville, AL 35812, USA}
\altaffiltext{5}{Dept. of Astronomy and Astrophysics, Pennsylvania State University, 525
Davey Laboratory, University Park, PA 16802, USA}
\altaffiltext{6}{Physics Dept., University of Alabama in Huntsville, 320
Sparkman Dr., Huntsville, AL 35805, USA }

%\documentclass[twocolumn]{aastex6}
%\usepackage{epsfig}

%\date{\today} % delete this line to display the current date
\def\ve{\varepsilon} \def\et3{\eta_3} \def\th1{\theta_{-1}} \def\r07{r_{0,7}}
\def\x05{x_{0.5}} \def\ve{\varepsilon} \def\muh{\hat{\mu}} \def\cm{\hbox{~cm}}
\def\km{\hbox{~km}} \def\kpc{\hbox{~kpc}} \def\Mpc{\hbox{~Mpc}}
\def\Hz{\hbox{~Hz}} \def\Gpc{\hbox{~Gpc}} \def\s{\hbox{~s}} \def\g{\hbox{~g}}
\def\gev{\hbox{~GeV}} \def\Jy{\hbox{~Jy}} \def\TeV{\hbox{~TeV}}
\def\GeV{\hbox{~GeV}} \def\MeV{\hbox{~MeV}} \def\kev{\hbox{~keV}}
\def\keV{\hbox{~keV}} \def\eV{\hbox{~eV}} \def\G{\hbox{~G}}
\def\erg{\hbox{~erg}} \def\s{{\hbox{~s}} \def\cm2{\hbox{~cm}^2}}
\def\para{\parallel} \def\Fl{\mathcal{F}} \def\tdec{t$_{\rm dec}$}
%\linenumbers

\begin{abstract} 
The possible short gamma-ray burst (GRB) observed by {\it Fermi}/GBM in
coincidence with the first gravitational wave (GW) detection, offers new ways
to test GRB prompt emission models.  Gravitational wave observations provide
previously unaccessible physical parameters for the black hole central engine
such as its horizon radius and rotation parameter. Using a minimum jet
launching radius from the Advanced LIGO measurement of GW~150914, we calculate
photospheric and internal shock models and find that they are marginally
inconsistent with the GBM data, but cannot be definitely ruled out.
Dissipative photosphere models, however have no problem explaining the
observations.  Based on the peak energy and the observed flux, we find that the
external shock model gives a natural explanation, suggesting a low interstellar
density ($\sim 10^{-3}$ cm$^{-3}$) and a high Lorentz factor ($\sim 2000$). We
only speculate on the exact nature of the system producing the gamma-rays, and
study the parameter space of a generic Blandford Znajek model. If future joint
observations confirm the GW-short GRB association we can provide similar but
more detailed tests for prompt emission models.  
\end{abstract}

\section{Introduction}
\label{sec:intro}
With the first detection of GWs we entered a new era in astrophysics
\citep{Abbott+16gw1}.  Electromagnetic counterparts are crucial for
establishing the astrophysical context for the GWs and also for a more accurate
localization to aid subsequent follow-up \citep{Connaughton+15loc}. GRB
progenitors \citep[see ][for reviews]{Meszaros+14review, Kumar+15review} have
been leading candidates for sources of GWs \citep{Kobayashi+03gwgrb,
Corsi+09mag}.  The most widely considered GW sources are compact binary mergers
with components stemming from a combination of neutron stars (NS) or black
holes (BH).  Other than BH-BH mergers, substantial radiation is expected to
accompany the GW signal, and indeed, the leading candidate for short-hard GRBs
are merging neutron stars \citep{paczynski86,eichler89}.  The GW~150914 event
is best explained by the merger of two $\sim$30 M$_\sun$ black holes.  {\it
Fermi}/GBM detected a tantalizing counterpart, GW~150914-GBM
\citep{Connaughton+16gbmgw}, consistent with a weak short GRB, broadly
consistent with the GW location and temporally coincident with the GW signal
(offset of $\Delta t_{\gamma-{\rm GW}} \approx t_{\rm GRB}-t_{\rm GW}$=0.4 s).
{We note however that while Advanced LIGO and GBM locations are consistent,
they both span a significant portion of the sky ($\sim$600 square degrees for
Advanced LIGO at 90\% confidence level  and  $\sim$3000  square degrees for GBM at
68\% confidence level).} This observation potentially marks the beginning of
the multi-messenger astrophysics.

In this paper we assume the weak GBM burst is a GRB (we refer to it as
GW~150914-GBM) associated with GW~150914 and investigate its implications for
the physical parameters of the system and for its surroundings.  This joint
electromagnetic (EM), GW observation was already addressed in a significant
number of early studies covering aspects of EM energy extraction from a binary
BH system and its surroundings \citep{Li+16gw, Loeb16gw,Perna+16gw,
Fraschetti16gw,  Yamazaki+16gw, Zhang16gw}. 

{INTEGRAL ACS observations \citep{Savchenko+16gwgrb} set a constraining
upper limit in terms of the source fluence in the ACS energy range (above
$\sim$75 keV).  The uncertainties on the GBM spectral parameters and on the
direction of the possible source, however, weaken any tension between the two
measurements \citep[for details, see Section 3.3 of ][]{Connaughton+16gbmgw}.}
{Nonetheless,} we emphasize that the association between the GBM event and
GW~150914 might have occurred by chance.  However, because the false alarm
probability of the two events being associated is P=$2.2\times10^{-3}$
\citep{Connaughton+16gbmgw}, we will {\it assume} a common origin and venture
to discuss the implications for the GRB emission models.

There has been considerable uncertainty in the GRB prompt emission model
parameters, such as the compact object mass and rotation rate. For the first
time however we can use realistic input parameters for modeling the BH central
engine, because the gravitational wave observations yield these parameters to a
precision that was previously unavailable. We calculate, to the extent that the
gamma-ray observations allow, the constraints on the usual GRB models that can
be placed.

Jets and black hole central engines are thought to be ubiquitous in GRBs.
Energy released from the central engine becomes collimated either by magnetic
stresses or the ram pressure of a progenitor star.  The initial dynamics of the
jet are determined by the launching radius, the size of the base of the jet,
where the Lorentz factor of the matter, which eventually produces the GRB, is
around unity. In other words the launching radius ($R_0$) is the characteristic
size of the volume in which energy is deposited. It is beyond this radius that
the jet starts to accelerate.  

Current methods of determining the launching radius rely on the blackbody
components in the GRB spectrum \citep{Peer+07lorentz}.  \citet{Larsson+15jet}
found the launching radius for GRB 101219B is approximately 10 times the
horizon radius.  This suggests the launching radius is defined by the scale of
the  BH central engine rather than larger scales ($\gtrsim 10^9 \cm$) such as
the progenitor star (e.g. in the case of reconfinement shocks,
\citet{Nalewajko11jetdiss}). Even considering substantial progress in jet
modeling, the launching radius is one of the least well constrained physical
parameters of the fireball model.  The gravitational wave observations can
determine the parameters of the resulting black hole and give a strict lower
limit on the launching radius.

In the next section we list the observational properties of the GW and
$\gamma$-ray event. In Section \ref{sec:model}, we briefly speculate on the
parameters of the gamma-ray emitting system.  In Section \ref{sec:rad}, we
mention GRB radiation models in the context of this source.  Finally, we
discuss our results in Section \ref{sec:disc}. For quantity Q, we use the
$Q_x=Q/10^x$ scaling notation in cgs units and the physical constants have the
usual meanings.

\section{Observations}
\label{sec:obs}
\subsection{Gravitational waves} 
The energy released in GW~150914, $E_{\rm GW}\approx 3 M_\sun c^2 \approx 5.4
\times 10^{54} \erg$ is comparable to the isotropic equivalent energy release
of very bright GRBs.  The final mass of the merged BH is $M_{\rm BH}=62\pm4
M_\sun$ and its rotation parameter is $a=0.67^{+0.05}_{-0.07}$
\citep{Abbott+16gw1}.  The gravitational radius of a 62 solar mass BH is $R_G=G
M_{\rm BH}/c^2 = 9.2\times10^6 \cm$, the horizon radius is
$R_H=(1+\sqrt{1-a^2})R_G=1.6\times 10^7 \cm$. 

The innermost stable orbit which we later associate with the launching radius
of the jet, is at $R_0 \approx f_1(a) R_G = 3.2 \times 10^7 \cm$.
$f_1(a=0.67)= 3.5$, where $f_1(a)=2-a+2(1-a)^{1/2}$.

GRB models usually invoke jets emitted along the rotation axis of their
progenitor BH. Due to Doppler boosting we can further assume our viewing angle
is within the opening angle of the jet, otherwise the EM emission would be
highly suppressed (essentially undetectable).  The most favorable configuration
for both GW detection and jetted emission is in case we view the binary system,
perpendicular to the rotation plane. The GW signal does not have a strong
dependence on the viewing angle. All things being equal, the difference in GW
signal-to-noise  from a face-on to an edge-on configuration is a factor of
$\sqrt{8}\approx2.8$.  Due to a known degeneracy between the inclination angle
and distance \citep{Cutler+94gw,Ligo16prop}, basically all the inclination
angles are allowed by GW data.  The observed $\gamma$-rays however suggest that
we see the system close to face-on.

\subsection{Gamma-rays}

\begin{figure*}[!htbp]
\begin{center}
\includegraphics[width=1.99\columnwidth,angle=0]{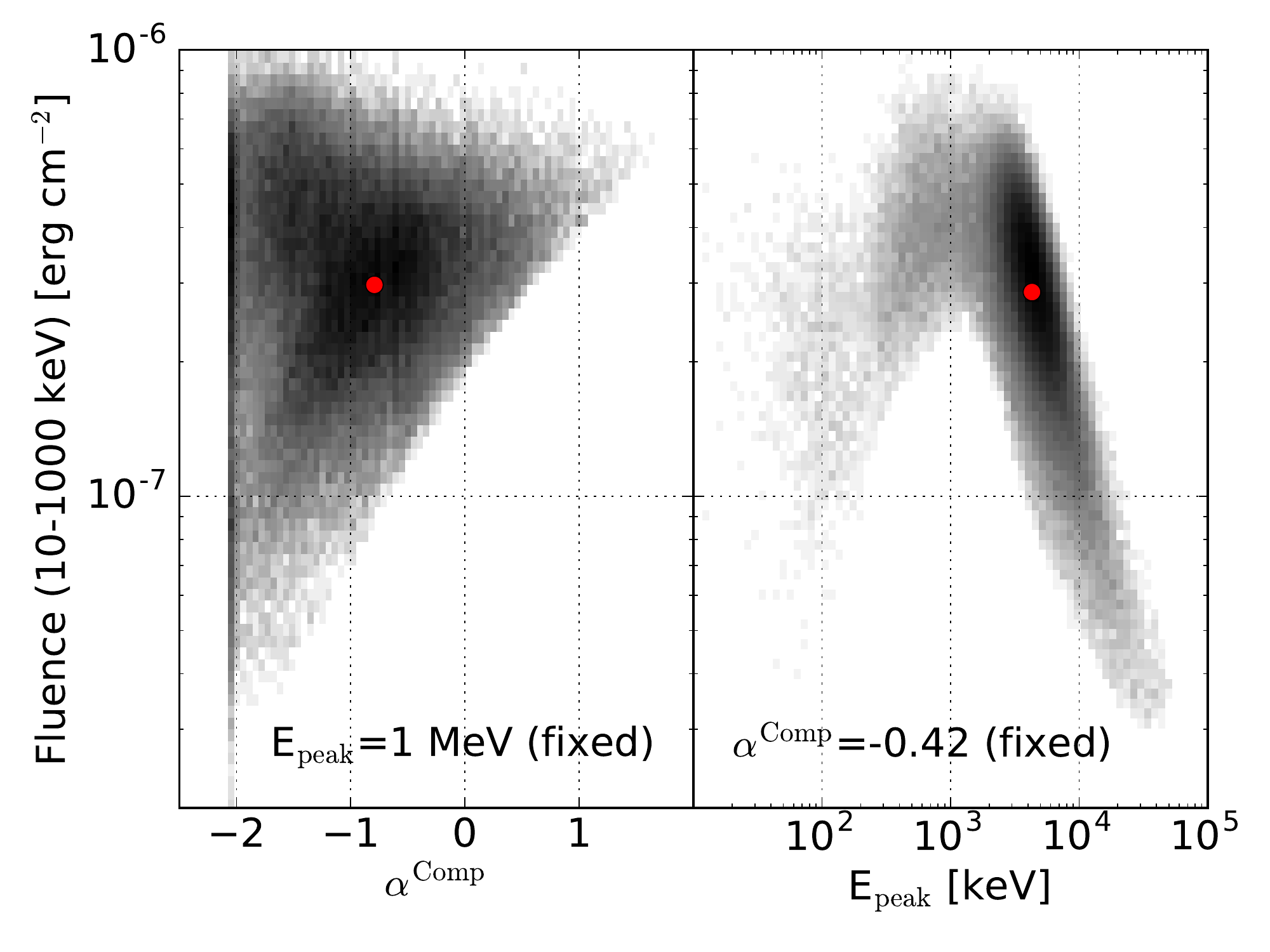}
\caption{Spectral parameters using the Comptonized model. In the left panel we
fix the peak energy to 1 MeV and display the distribution of the power law
index versus fluence. In the right panel we plot the distribution of the peak
energy similarly against fluence, when the power law index is fixed at the
average for short GRBs, $\alpha^{\rm Comp}=-0.42$. Red dots mark the peak of
the histograms.
} \label{fig:sp1} \end{center} \end{figure*}

GBM  detected a weak source with duration of $T_\gamma\approx 1 \s$ which, at
the time of the GW trigger was in an unfavorable position for the GBM
detectors.  Careful analysis however reveals a source with parameters
consistent with a short GRB \citep{Connaughton+16gbmgw}. 

GBM count spectra were deconvolved with detector response matrices for multiple
positions within the joint GBM/LIGO localization region. The resulting fit is
mostly consistent with a hard power law, with photon index $\alpha_{\rm
PL}=-1.40^{+0.18}_{-0.24}$.  This power law index is consistent with other weak
short GRBs that have been detected by GBM that can only be fit by a power law
due to their low flux.  The median (mean) value for the PL index for all weak
GBM short GRBs is -1.36 (-1.40).  The luminosity in the 1 keV to 10 MeV range,
which is a good approximation for the bolometric luminosity, is $L_{\rm
obs}\approx 2.3\times 10^{49}$ erg s$^{-1}$.

Hints for a possible cutoff energy come from one of the positions on the
initial localization annulus, where it was possible to constrain a physically
more realistic, Comptonized spectrum (power law with exponential
cutoff):\footnote{We note that while the cutoff energy could be constrained,
the cutoff power law model is not statistically preferred over a power law
spectrum.}

\begin{equation} 
\frac{dN_{\rm ph}}{dE} = A \left(\frac{E}{100\,{\rm
keV}}\right)^{\alpha^{\rm Comp}} \exp\left({ -\frac{E (\alpha+ 2)}{E_{\rm pk}^{\rm
Comp}}}\right).  
\label{eq:cpl}
\end{equation}

Short GRBs are typically well fit using this model, with high peak energies and
a steep spectrum above the peak energy.    The assumed source position however
is not compatible with the relative count rates measured in the 14 GBM
detectors, and is excluded at the 90\% confidence level by the {\it joint}
GBM-LIGO localization.

The event is too weak to confidently constrain more than two spectral
parameters.  Since the best fitting power law spectrum is ultimately unphysical
because it implies infinite amount of liberated energy, we carry out Monte
Carlo simulations to evaluate the ranges of the Comptonized model parameters
allowed by the observations.  The count spectrum has a maximum at roughly 1
MeV, indicating that the spectral peak lies at that energy or above it,
regardless of the details of spectral fitting. We substantiate this claim by
fixing the photon index at the mean value for short GRBs, -0.42 for a model
with an exponential cut-off above $E_{\rm pk}^{\rm Comp}$, and generate a
distribution of amplidudes and peak energies for all the positions along the
localization arc. We find that in 94.6\% of the cases the peak energy exceeded
1 MeV indicating a high peak energy event similar to short GRBs (Figure
\ref{fig:sp1}, right). We also carry out a simluation where we fix the peak
energy at 1 MeV and fit the amplitude and the photon index (Figure
\ref{fig:sp1}, left).

\section{Radiation from a stellar mass black hole merger}
\label{sec:model}

Significant EM energy release from a binary BH merger is unexpected.  The
dispersion length of gravitational waves is much larger than the curvature
radius associated with material surrounding the merger in any conceivable
scenario \citep[e.g.][]{MTW}, thus no significant energy transfer is expected
from GW to matter.  Also, no obvious debris are expected from the BH-BH
merger that can aid the energy release, similar to a BH-NS or NS-NS mergers
\citep[see however][]{Loeb16gw, Murase+16gw, Perna+16gw, Yamazaki+16gw,
Li+16gw}.

In principle, a fraction $f_r(a)=1-\sqrt{(1+\sqrt{1-a^2})/2}\approx6\%$ of the
BH energy is available for extraction from a rotating BH. This corresponds to
$E= 62 f_r(0.67) M_\sun c^2\approx6.7\times 10^{54}\erg$ which is four orders
of magnitude more than the observed energy release.  Methods of tapping energy
from the merged BH include neutrino driven disks \citep{Zalamea+11nu},
Blandford-Znajek (BZ) process \citep{Blandford+77znajek} (see however
\citep{Lyutikov16gw}).

More conventional models for energy release in a BH system can be put forward
as follows: In the GW~150914 progenitor system about $5\%$ of the initial total
mass was radiated away as GW. Because of the reduced central mass, orbits of
the fluid elements in a disk around the final BH will be modified
\citep{Bode+07circ}, possibly producing shocks. Furthermore, the final BH will
experience a kick associated with the anisotropic emission of GWs
\citep{Farris+11bhmerger}.  The angular momentum vectors of the binaries were
parallelized by interaction on long time scales with circumstellar matter
implying the kick will launch the black hole into the surrounding disk with
$v\lesssim 1000 $ km s$^{-1}$, possibly enhancing the accretion rate. However,
these mechanisms yield only $\lesssim L_E$ luminosities for conditions normally
expected around BH mergers \citep[e.g.][]{Lippai+08shock}. 

\subsection{Generic Blandford-Znajek scenario}
We forgo pursuing the exact nature of energy release from the BH merger which
later results in the electromagnetic counterpart. Specifically, we do not
address the provenance of the disk material required to tap the energy of the BH.
Given the unexpected association of the GW and EM signals, we outline the
generic properties of a BH-disk system launching a jet with opening angle
$\theta_{\rm jet}=0.1$ assuming the BZ mechanism. Our aim is to constrain the
parameter space for this particular scenario through specific criteria.

\begin{figure*}[!htbp]
\begin{center}
\includegraphics[width=1.99\columnwidth,angle=0]{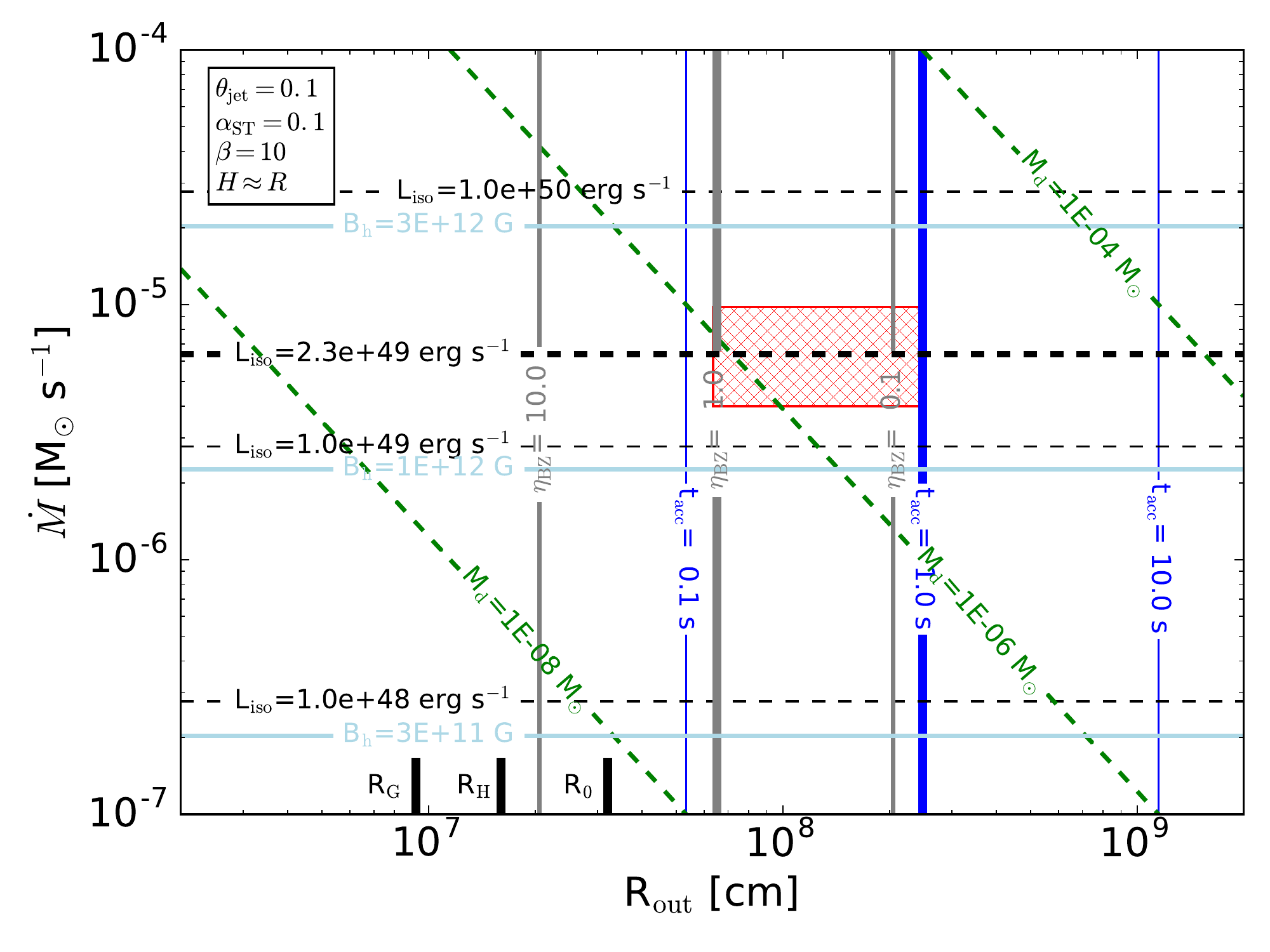} 
\caption{Disk parameter space constrained by the observations. The cross-hatched
rectangle marks the approximate allowed outer disk radius and mass accretion
rate.  This is valid for an assumed jet opening angle of $\theta_{\rm
jet}=0.1$.  Isotropic equivalent (observed) luminosity is in black dashed
lines, dark blue lines show the accretion timescale, light blue lines show the
magnetic field strength, green dashed lines show the disk mass and gray lines
show the BZ efficiency factor.  On the bottom x axis we indicate the values of the
gravitational radius, the horizon radius and the marginally stable (innermost
stable) radius.} \label{fig:b1} \end{center} \end{figure*}

We assume the disk height is $H(R)=R$, but for $H(R)\approx 0.3 R$ as required
by the model of \citet{Perna+16gw} the allowed parameter space does not change
significantly.  We illustrate the system parameters on a $R_{\rm disk}-\dot{M}$
plane (see Figure \ref{fig:b1}).  The disk has a viscosity parameter
$\alpha_{\rm ST}=0.1$ \citep{Shakura+73disk}.

As a first criterion, we turn to the timescales governing the launch of the
putative jet. The accretion timescale, during which a jet of outer radius $R_{\rm
out}$ is swallowed by a BH, can be calculated from the viscous timescale of the
disk. As an example, for $R_{\rm out} \sim 2\times 10^8 \cm$, the accretion
time is $t_{\rm acc}= (7/3\alpha) \sqrt{R_{\rm out}^3 /G M_{\rm BH}} \approx 1
~\alpha_{-1}^{-1} (R_{\rm out}/2\times10^{8}\cm)^{3/2}~ (M_{BH}/62
M_\sun)^{-1/2}{\s} $.  The central engine timescale (the accretion timescale in
our example, \citet{Zhang+09typeI}) can be constrained to be at most the
observed duration or $t_{\rm acc}\lesssim 1 $ s \citep{aloy05}. This constraint
will carve out a region in Figure \ref{fig:b1} to the left of the $t_{\rm acc}=
1 $ s or $R_{\rm out}=2.3\times 10^8 \cm$ line.

Another criterion which can constrain the physical parameters of the system is
the activation of the BZ mechanism. In order for the BZ mechanism to occur, the
magnetohydrodynamic waves need to be able to escape the BH ergosphere
\citep{Komissarov+09BZ}. To first approximation, this translates to the
Alfv\'en velocity, $v_A^2\approx B^2/4\pi \rho$, being larger than the
free-fall velocity, $v_{\rm ff}^2=2GM_{\rm BH}/R$.  Here $\rho$ is the disk
density.  Defining $\eta_{\rm BZ}=v_{\rm ff}/v_A = 4 \pi \rho c^2 / B^2$, for
efficient energy extraction we require $\eta_{\rm BZ}<1$. The exact limit,
depending on the rotation parameter, can be obtained by numerical simulations
\citep{Komissarov+09BZ} and for $a=0.7$ (see their Figure 3) it can go as low
as  $\eta_{\rm BZ}<0.3-0.5$.  We conservatively take $\eta_{\rm BZ}<1$ at
$R\approx R_H$, to result in a constraint of $R_{\rm out}\gtrsim 7\times 10^7
\cm$ (see Figure \ref{fig:b1}).

The accretion velocity (for adiabatic index $\hat{\gamma}=4/3$) can be written
as $v_{\rm in}=(3\alpha/7) v_k$. The sound speed is $c_s^2=(2/7) v_k^2$, where
$v_k=\sqrt{G M_{\rm BH}/R}$ is the Kepler velocity.  Writing the mass accretion
rate as, $\dot{M}=2\pi R H \rho v_{\rm in}$, we can express the pressure as
$\hat{\gamma}P=\rho c_s^2$.

Assuming the magnetic field threading the BH at the marginally stable radius
($R_0$) is related to the thermal pressure through the magnetization parameter
$\beta$, ($P=B^2\beta/(8\pi)$) according to \citet{Komissarov+10pop3}, we can
write:
\begin{equation}
B\approx5.4 \times 10^{11} \dot{M}_{-5}^{1/2}
\left(\frac{f_1(a)}{2.48}\right)^{-5/4} \left(\frac{M_{\rm BH}}{62
M_\sun}\right)^{-1} \alpha_{-1}^{-1/2} \beta_{1}^{-1/2} \G. 
\end{equation}
where $f_1(a=0.67)=2.48$.

Next, we consider the BZ luminosity ($L_{\rm BZ}$).  We define $\Psi_h=2\pi R_h^2
B_h$ as the magnetic flux threading the BH horizon, $\Omega_h=f_2(a) c^3/G
M_{\rm BH}$ is the angular velocity and  $f_2(a)=a/(2+2\sqrt{1-a^2})$. 
The expression for the BZ luminosity:
\begin{eqnarray}
L_{\rm BZ}&=&\frac{1}{3c} \left(\frac{\Psi_h \Omega_h}{4\pi} \right)^2\nonumber \\
&=&\frac{\sqrt{14} f_1^{3/2}(a) f_2^2(a)}{9 \alpha \beta} \dot{M}c^2\nonumber \\ 
&=&1.1\times 10^{48} \frac{\dot{M}_{-5}}{\alpha_{-1}\beta_{1}} \erg \s^{-1}. 
\label{eq:LBZ}
\end{eqnarray}

If the radiation emanates from a jet with half opening angle, $\theta_{\rm
jet}$, the observed, isotropic equivalent luminosity is related to the BZ
output power as $L_{\rm iso} = L_{\rm BZ} \theta^{-2}_{\rm jet}/2   = 200
(\theta_{\rm jet}/0.1)^{-2} L_{\rm BZ}$.  Thus, the observed isotropic
equivalent luminosity can be expressed as a function of $\dot{M}$ (assuming a
jet opening angle) and it is possible to use it to  constrain the allowed
parameter space of $\dot{M}$ (see $L_{\rm iso}$ lines on Figure \ref{fig:b1}).
We draw the hatched region by allowing a factor of $\sim 2$ around the observed
luminosity of $L_{\rm iso}=2.3\times10^{49} $ erg s$^{-1}$.  In this scenario,
we can constrain the disk mass to $10^{-6} - 10^{-5} M_{\odot}$, and the
magnetic field threading the horizon will be  $\sim 2\times10^{12} \G$.

%%%%%%%%%%%%%%%%%%%%%%%%%%%%%%%%%%%%%%%%%%%
\section{Radiation mechanisms} 
\label{sec:rad} 
Focusing on the gamma-ray signal, it has both spectral and temporal properties
consistent with the prompt emission of a short GRB.  The observed luminosity is
$\sim$10 orders of magnitude larger than the Eddington luminosity, indicating
gamma-ray source has likely experienced relativistic expansion.  In the GRB
fireball scenario the jet Lorentz factor is expected to go through the
acceleration phase, starting from the launching radius ($R_0$), $\Gamma(R)=
R/R_0$.  When the accelerating material has kinetic energy per unit particle of
$\sim \eta m_p c^2$ where $\eta =L/\dot{M}c^2 (\gtrsim100$ for GRBs) is the
dimensionless entropy, we have $\Gamma\approx\eta=$constant. This point is
marked by the saturation radius, $R_{\rm sat}=R_0\eta$.  Starting at the
deceleration radius (see Section \ref{sec:ES}), $R_{\rm dec}$, the jet enters
the self-similar deceleration phase, where $\Gamma(R) =\eta (R/R_{\rm
dec})^{-g}$, $3/2<g<3$ \citep{Meszaros+93gasdyn}.  Because of the weakness of
the $\gamma$-ray source, we will only address peak energies of models and in
some cases their fluence.

\subsection{Photospheric models} 

In the relativistically expanding material the location where the Thompson
scattering optical depth falls below unity marks the position of the
photosphere.  This is the innermost radius from where radiation can escape and
can be calculated from $R_{\rm phot}= L\sigma_T/8\pi m_p c^3
\Gamma_{\rm phot}^2 \eta$:
\begin{widetext}
\begin{equation}
R_{\rm phot}\approx \left\{
\begin{array}{lll}
    2.8\times 10^{7} ~  \left(\frac{L}{L_{\rm obs}}\right)\eta_3^{-3} \cm
		& {\rm if~ }  R_{\rm phot}>R_{\rm sat}& {\rm or~ } \eta<\Gamma_T \\
    4.6\times 10^{9} ~   \left(\frac{L}{L_{\rm obs}}\right)^{1/3} 
		\left(\frac{R_0}{R_*}\right)^{2/3} \eta_3^{-1/3} \cm
		& {\rm if~ }  R_{\rm phot}<R_{\rm sat} & {\rm or~ } \eta>\Gamma_T, 
	\end{array}
	\right. 
\end{equation}
\end{widetext}
where $\Gamma_T=170~(L/L_{\rm obs})^{1/4} (R_0/R_*)^{-1/4}$ is the Lorentz
factor separating the photosphere in the acceleration phase ($\eta>\Gamma_T$)
and the photosphere in the coasting phase ($\eta<\Gamma_T$). We note here
that based on the joint GW and EM observations this quantity can be well
determined.  Henceforth, for brevity we use $L_{\rm obs}=2.3\times 10^{49} \erg
\s^{-1}$ and $R_{*}=3.2\times10^7 \cm$ to mark the numerical scaling values for
the observed luminosity and the calculated launching radius respectively.
$R_{\rm sat}=R_0 \Gamma\approx 3.2\times 10^{10} ~(R_0/R_*)~ \Gamma_3 \cm$. The
fact that the actual Lorentz factor, $\Gamma \gtrsim 1000$ is likely larger
than $\Gamma_T$ (see e.g. Section \ref{sec:ES}) suggests the photosphere occurs
in the acceleration region. E.g. for $\Gamma=10^3$, we have $R_{\rm phot}\approx
3\times 10^9 \cm <R_{\rm sat}$.
The observed temperature of an expanding fireball at its
photosphere (occurring in the acceleration phase) can be calculated as
$T_0\approx (L /4\pi R_0^2 c a)^{1/4}\approx 1 (L/L_{\rm obs})^{1/4}
(R_0/R_*)^{-1/2}\MeV$.  The maximum attainable peak energy of a spectrum with
temperature $T$ \citep{Li+08thermal,Fan+12corr,Zhang+12epeak}:
\begin{equation}
E_{\rm pk}^{\rm PH}\lesssim 3.92 \times kT_0 \approx  0.6  \left(\frac{L}{L_{\rm
obs}}\right)^{1/4} \left(\frac{R_0}{R_*}\right)^{-1/2} \MeV.
\label{eq:phpk}
\end{equation}
The 3.92 factor indicates that the $E_{\rm pk}$ is the peak in the $\nu F_\nu$
representation.  The fact that we know $R_*$, the smallest possible launching
radius from the GW observations, we know Equation \ref{eq:phpk} is a strict
upper limit for the temperature of the non-dissipative photosphere.  The
measured peak can reach the upper limit in Equation \ref{eq:phpk} in the case
where the photosphere occurs in the acceleration region. In this case the
comoving temperature is proportional to R$^{-1}$ and the increase of the
Lorentz factor ($\propto R$) compensates the decrease to yield a temperature of
$T_0$. 

The peak energies of the simulated set of Comptonized spectra violate this limit
(see Figure \ref{fig:sp1}, right) in an overwhelming number of cases.  Thus,
even with the uncertainties of the spectral parameters, we consider this model
is not favored by the data, however we cannot rule it out.

A more sophisticated class of models for the GRB prompt emission are {\it
dissipative} photosphere models \citep{Rees+05photdis}.  In these models energy
is liberated while the flow is still optically thick through e.g.
neutron-proton collisions \citep{Beloborodov10phot} or magnetic reconnection
\citep{Giannios+07photspec}.  There are no simple criteria for meaningful
comparisons with data for GW~150914-GBM.  For these models, in general terms,
the peak energy is not constrained by the expression in Equation \ref{eq:phpk}
and can reach substantially higher values $\lesssim 20 \MeV$
\citep{Beloborodov12epeak}. E.g. for the observed luminosity of $2.3\times
10^{49} \erg \s^{-1}$, the maximum achievable peak energy is around $\sim 10
\MeV$ both for magnetic field dominated outflows and for baryon dominated cases
 as well \citep[see Figure 2 of ][]{Veres+12peak}.

\subsection{Internal shocks} 
Internal shocks can occur in unsteady relativistic outflows \citep{Rees+94is}
where a faster shell catches up with a slower one.  The colliding shells
 produce shocks that accelerate electrons, amplify the magnetic field and in
turn the electrons emit synchrotron radiation. This process can
tap the relative kinetic energy of material ejected at different times from the
central object. 

The radius of  internal shocks can be calculated as $R_{\rm IS}\approx 2
c\Gamma^2 dt$, where $dt$ is the variability timescale. For short GRBs, the
average $dt$ is $\approx10^{-2} \s$ \citep{MacLachlan+13tvar}.  The detailed
temporal structure of the gamma-ray lightcurve could not be determined by the
GBM data for this weak event.  Based on the GW observations, however, we can
put a lower limit on the variability timescale that is the dynamic time $t_{\rm
dyn}=R_0/c=1.1\times 10^{-3}~ R_0/R_* \s$ such that $dt\gtrsim t_{\rm dyn}$.
The internal shock radius ($R_{\rm IS}$) will be well above the photosphere,
implying an optically thin outflow.  

The synchrotron peak frequency will be at $E_{\rm peak} =  h q_e/(2\pi m_e c)
\Gamma \gamma_e^2 B$.  For $\gamma_e=(m_p/2m_e) \epsilon_e\approx 940~\epsilon_e$
and  $B=\sqrt{2 \epsilon_B L T_\gamma/(\Gamma^2 c dt)^3}=4\times 10^7 ~(L/L_{\rm
obs})^{1/2} \epsilon_B dt_{-3}^{-3/2} T_{\gamma,0}^{1/2} \G $
\citep{panaitescu00b}, the peak will be at:
\begin{equation}
E_{\rm pk}^{\rm IS}\lesssim 0.54  ~(L/L_{\rm obs})^{1/2} \Gamma^{-1}_3 dt_{-3}^{-3/2}
\epsilon_B^{1/2} \epsilon_e^2 T_\gamma^{1/2}\MeV. 
\end{equation}
A more thorough analysis considering pairs and the width of the shells
\citep{Guetta+01} yields a stricter limit:
\begin{equation}
E_{\rm pk}^{\rm IS}\lesssim 80 ~ (L/L_{\rm obs})^{1/6} (\Delta/R_0)^{-5/6}
dt_{-3}^{1/6} \epsilon_B^{1/2} \epsilon_e^{4/3} \keV. 
\end{equation}
Here, $\Delta$ is the width of the shell which has to be greater than $R_0$.
Note that here, the electron and magnetic field equipartition parameters
$\epsilon_e$ and $\epsilon_B$ are normalized to 1.

With the  spectral peak constrained to be reliably above 1 MeV, the above
derivation suggests the internal shock model has difficulties in accounting for
peak energies above 0.5 MeV with a given launching radius.  Keeping in mind the
large errors on the observed quantities, we can say this model does not
naturally explain the observed peak energy, however, just as in the case of the
non-dissipative photosphere models, we cannot completely rule it out.

\subsection{External shocks} 
\label{sec:ES}

External shocks were initially proposed \citep{Rees+92fball} as a model for GRB
prompt emission, but had problems interpreting the strong variability of
lightcurves \citep{kobayashi97} (see however \citet{Dermer+99es}).  On the
other hand is a very successful model for interpreting the multiwavelength
afterglow \citep{Meszaros+97ag,chiangdermer99}.  Recently however, claims of
external shock origin for the prompt emission have been reported for bursts
with smooth, simple lightcurves \citep{Burgess+16es}.  External shocks are
almost guaranteed to form around a relativistically expanding shell. Here we
apply the formalism of the external shock model to constrain the physics of the
GW associated  GW~150914-GBM event.

A shock front develops as the material from the central engine interacts with
the interstellar material (ISM).  The timescale on which this occurs is the
deceleration time. It marks the time where the material plowed up by the
relativistic jet corresponds roughly $1/\Gamma$ times the mass in the ejecta.
For interstellar material of number density $n$, Lorentz factor $\Gamma$, and
kinetic energy $E_k$ we have
\begin{equation}
t_{\rm dec} \approx 0.28~ n_{-3}^{-1/3}\Gamma_{3}^{-8/3}
(E/E_{k})^{1/3} ~ {\rm s}.
\end{equation}
Based on \citet{zhang07a}'s results for short GRBs,  we assume a radiative
efficiency $\eta_\gamma=E_{\rm ph}/(E_{\rm k}+E_{\rm ph})=0.5$ and get
$E_k\approx E_{\rm ph}\approx L_{\rm obs} \times 1 \s\approx 2.3\times
10^{49} \erg$.

The peak flux density of the spectrum can be calculated by adding the
individual electron
powers and according to \citep{Sari+98spectra,Gao+13syn}:
\begin{equation}
    F_{\nu,{\rm p}}=\frac{N_e P_{\nu,\max}}{4\pi D_L^2},
\end{equation}
where $P_{\nu,{\rm max}}=m_e c^2 \sigma_T \Gamma B /3 q_e$ is the single
electron synchrotron power, $N_e=4\pi R_{\rm dec}^3 n/3$ is the number of swept
up electrons, $R_{\rm dec}\approx 2\Gamma^2 c t_{\rm dec}$ is the deceleration
radius and $B=\sqrt{32\pi \epsilon_B n m_p c^2 \Gamma^2}$ is the magnetic field
in the shocked region. $\epsilon_e$ and $\epsilon_B$ are the electron and
magnetic field equipartition parameters respectively.

\begin{figure*}[!tbp]
\begin{center}
\includegraphics[width=1.99\columnwidth,angle=0]{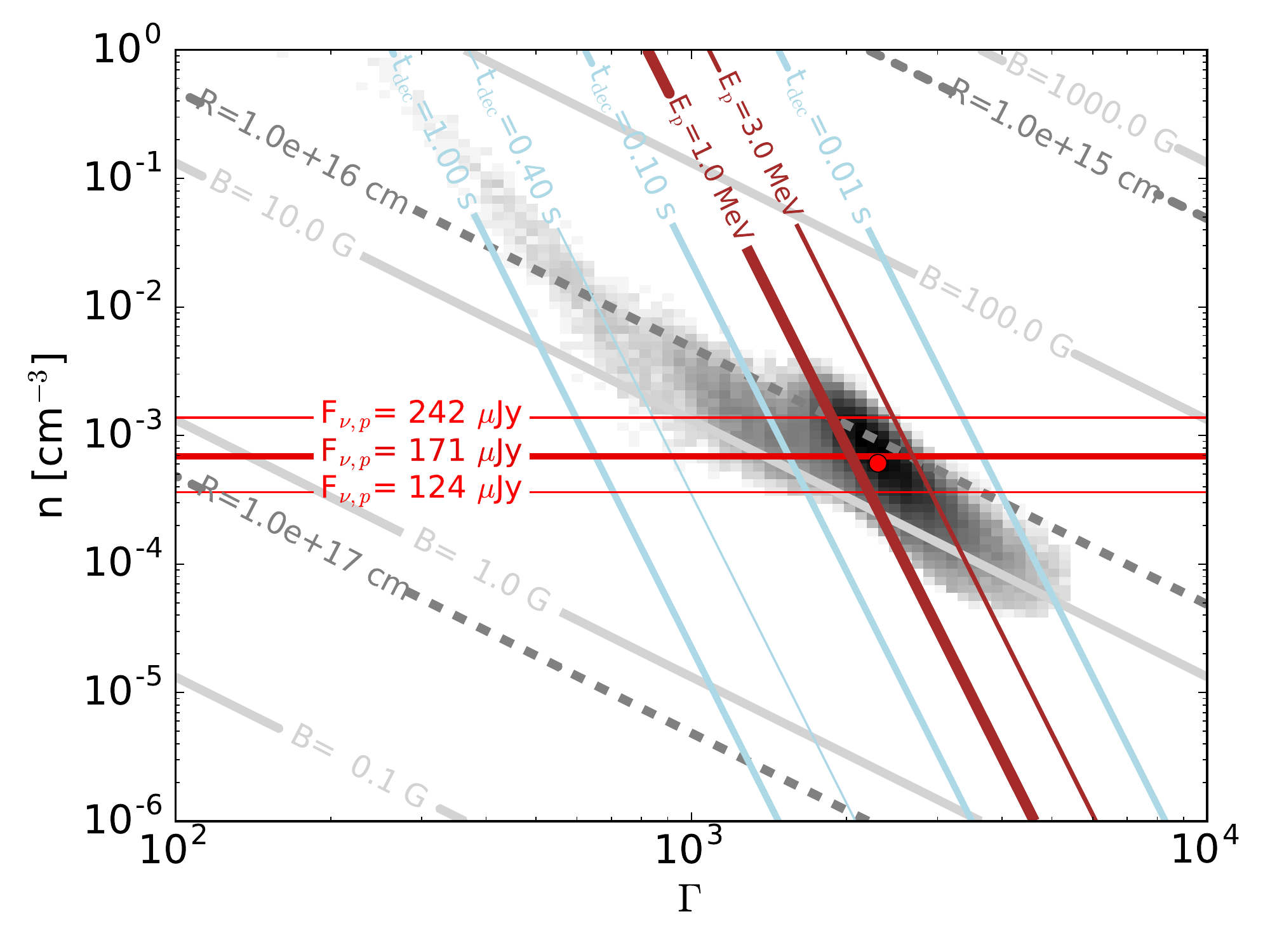} 
\caption{Lorentz factor- ISM density plane with constraints plotted for the
external shock scenario. Here, the photon index in the Comptonized model is
fixed to $-0.42$. Different colors and styles mark different physical
quantities, all displayed on the contour lines. The intersecting regions of 
brown and red contour lines ($E_{\rm peak}$ and $F_{\nu,{\rm peak}}$) mark the
favored parameter space (the red circle is at $n=8\times 10^{-4} \cm^{-3}$ and
$\Gamma=2300$). The underlying histogram shows the results of a Monte Carlo
simulation for the allowable spectral parameters. Darker bins indicate more cases.
We have assumed $\epsilon_e=\epsilon_B=0.5$. }
\label{fig:b2}
\end{center}
\end{figure*}

The peak energy, corresponding to electron random Lorentz factor of
$\gamma\approx 600\epsilon_e \Gamma$ is 
\begin{equation}
E_{\rm pk}^{\rm ES}=\frac{q_e B \gamma^2 \Gamma}{2 \pi m_e c}=
1.1 
\left(\frac{\Gamma}{2000}\right)^4
n_{-3}^{1/2}
\left(\frac{\epsilon_e}{0.5}\right)^2
\left(\frac{\epsilon_B}{0.5}\right)^{1/2}
\MeV.
\end{equation}
The peak of the external shock radiation occurs approximately at the
deceleration time. We know the delay of the EM trigger compared to the GW
signal, and we require $t_{\rm dec}\lesssim \Delta t_{\gamma-{\rm GW}}$.  We
draw the deceleration time values on Figure \ref{fig:b2} with light blue, and
note that our Monte Carlo simulated spectral parameters overwhelmingly result
in a deceleration time lower than the $\Delta t_{\gamma-{\rm GW}}$.

We use the measurement of the peak energy (the fact that it is likely above 1
MeV)  and the peak flux density from the Comptonized spectrum in the case for
the fixed photon index (Figure \ref{fig:sp1}, right) to place constraints on
the particle number density around the progenitor and the Lorentz factor of the
outflow. On Figure \ref{fig:b2} we show that the two constraints mark a region
centered around $n\approx6\times 10^{-4} \cm^{-3}$ and $\Gamma \approx 2300$.
The shaded region on Figure \ref{fig:b2} shows a two dimensional histogram of
simulated values (darker shades mark more cases within the region)
pointing out the non-trivial shape of the uncertainties on $\Gamma$ and $n$.
The peak flux lines (red) are the median and the values corresponding to the
full width at half maximum of the $F_{\nu,{\rm p}}$ distribution.  The
corresponding magnetic field strength is $B\approx 20 \G$, the radius of peak
emission is $R_{\rm dec}\approx 10^{16} \cm$ and the deceleration time is
$t_{\rm dec}\approx 4\times 10^{-2} \s$. With these parameters the synchrotron
radiating electrons are in the fast cooling regime, which means they lose their
energy faster than the dynamical timescale and this is in line with
expectations for the prompt emission.  This exercise can be carried out with a
wind profile interstellar medium, but for compact mergers the constant density
profile is preferred \citep[e.g.][]{panaitescu06b}.

{We note here that in the external shock scenario in its simplest,
impulsive energy injection case, the timescale of the GRB duration is also
governed by the deceleration time.  However, since the derived t$_{\rm dec}$
($4\times 10^{-2}\s$) and the GRB duration ($T_{\rm GRB}\approx 1 \s$) differ,
we argue that the GRB duration reflects the energy injection timescale, which
can be of the order of 1 s, instead of t$_{\rm dec}$.  }

{We set the microphysical parameters ($\epsilon_e$, $\epsilon_B$) to $0.5$.
Although we are using this model to constrain the prompt emission, based on
afterglow modeling these parameters would in general have lower values. E.g.
typical afterglow-based values would be $\epsilon_e=0.1$ and
$\epsilon_B=10^{-2}$.  The decrease in the microphysical parameters would have
to be compensated by an increase of the Lorentz factor and ISM density (to have
the same peak energy and flux) which would result in $\Gamma \sim 5300$ and
$n\sim2.2\times 10^{-2} \cm^{-3}$.  }

\subsubsection{Efficiency of external shocks}
The above results were presented for an efficiency of 0.5 ($E_k=E_{\rm ph}$).
In the external shock scenario, the radiation taps the kinetic energy of the
explosion. Higher than 50\% radiative efficiency is unexpected for external
shocks.  \citet{Dermer+99es}, for example, find that  10\% ($E_k=9E_{\rm ph}$)
efficiency is more consistent with this scenario.  In this case, the measured
fluence and the peak energy yields a Lorentz factor and external density of
($4\times 10^{-5} \cm^{-3}, 3300$). These are even more extreme than the values
for 50\% efficiency, but consistently point to the low density origin of the
binary source. The deceleration time is somewhat larger in this case, $t_{\rm
dec}\approx 8\times 10^{-2} \s$, but still within the $t_{\rm GRB}-t_{\rm
GW}$=0.4 s.

\subsubsection{External shock model with an average short GRB}

\begin{figure*}[!htbp]
\begin{center}
\includegraphics[width=1.49\columnwidth,angle=0]{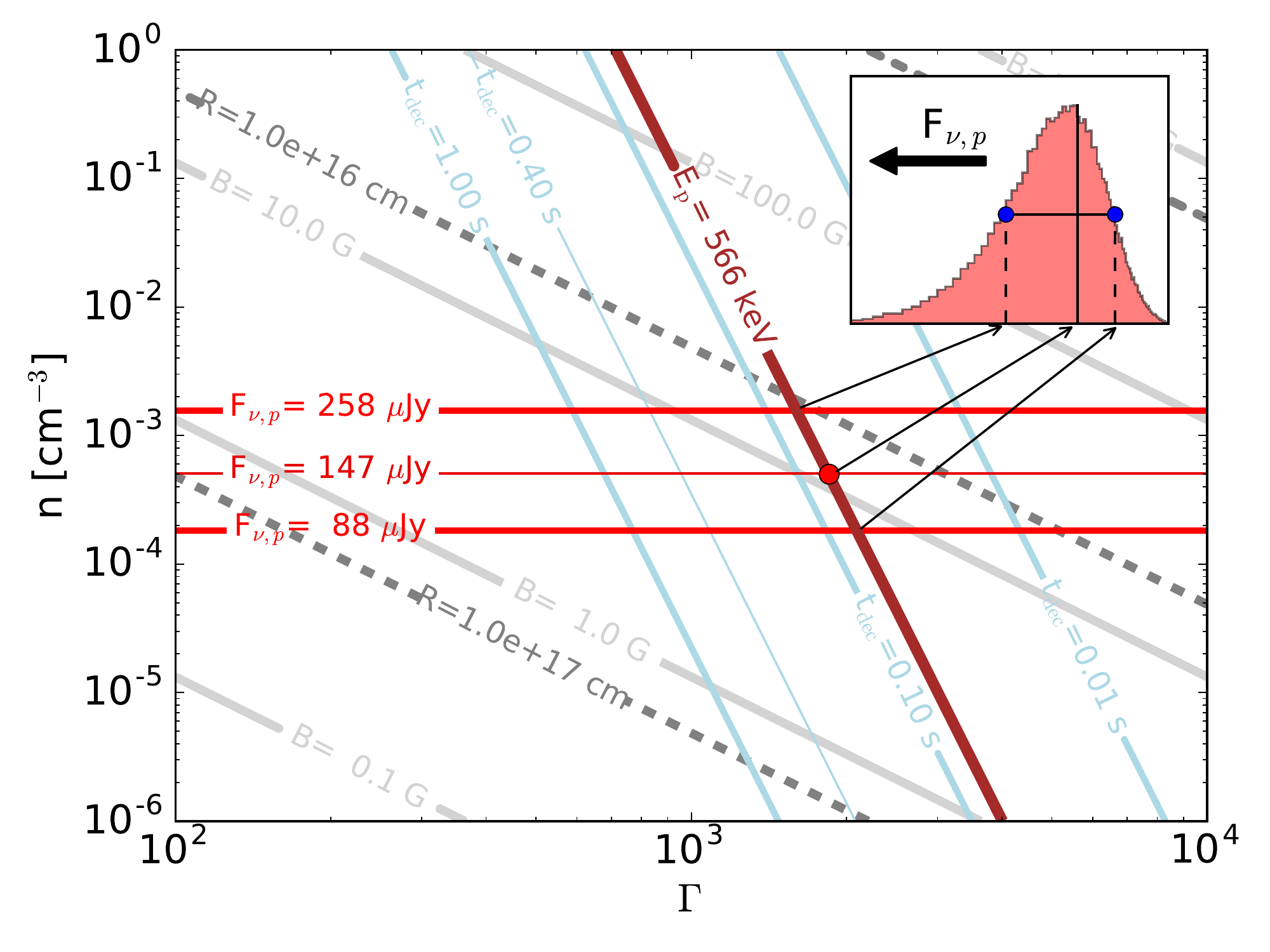} 
\caption{Lorentz factor- ISM density plane with the peak energy fixed to the
average value for short GRBs, and peak flux results from spectral fitting.  The
contours are marked similar to Figure
\ref{fig:b2}  with the physical quantities indicated on the contour lines.  The
inset shows the peak flux distribution with the thin arrows indicating the
median value and the errors in the distribution. The thick arrow indicates
F$_{\nu,p}$ values increase to the left. The red circle marks the most likely
set of parameters, $n=5\times 10^{-4} \cm^{-3}$ and $\Gamma=1800$. As before,
we have assumed $\epsilon_e=\epsilon_B=0.5$. }
\label{fig:b2s}
\end{center}
\end{figure*}

Because the peak energy is difficult to constrain for this event we analyze
another possible scenario. We take the {\it average} photon index and peak
energy for a Comptonized spectrum of the short GRB sample for GBM
\citep{Gruber+14cat}\footnote{ The up-to-date catalog is located at:
\url{http://heasarc.gsfc.nasa.gov/W3Browse/fermi/fermigbrst.html}}.  We fit the
spectral data and generate amplitude parameters ($A$ in Equation \ref{eq:cpl})
for the Comptonized spectrum by fixing $\alpha_{\rm short}=-0.42$, $E_{\rm
pk,short}^{\rm Comp}=566 \keV$ and using response matrices generated for the 11
positions along the Advanced LIGO localization arc.  We get a distribution of
peak fluxes for the fixed value of the peak energy. It is possible then to put
these parameters to the Lorentz factor density plane (see Figure \ref{fig:b2s})
and deduce that the required density distribution peaks around $5\times 10^{-4}
\cm^{-3}$ with a tail extending to a few$\times 10^{-3} \cm^{-3}$ and the
Lorentz factor is $\Gamma\approx1800$. The deceleration time for this case is
$\approx$ 0.1 s, which is again consistently lower than the 0.4 s delay between
the GW and EM signal.

\section{Discussion and Conclusion}
\label{sec:disc}

Assuming the binary black hole merger is associated with the gamma-ray signal
detected by GBM, we have taken the leading prompt emission models for GRBs and
applied it to the observations of GW~150914-GBM, aided by the accurately
determined central engine parameters through GW measurements.  We find that the
non-dissipative photosphere and the internal shock models have some difficulty
in interpreting the observations, though at this point no model can be
definitely ruled out, while a dissipative photosphere model is unconstrained.
The external shock model is able to interpret both the high peak energy and the
observed flux yielding  constraints on the Lorentz factor of the explosion
($\gtrsim 1000$) and the interstellar density ($\sim 10^{-3} \cm^{-3}$).  The
lower than usual ISM density is in line with the expectation that the merger
takes place far from the birthplaces of its components (e.g. in a galactic halo
environment). Furthermore, the  low density might be a more general property of
the external shock model itself which, applied to model afterglow observations
of  GRBs with $\gtrsim$ GeV photons (e.g. GRB 090510A) yield similarly high
$\Gamma$ and low $n$ \citep{DePasquale+09-090510ag}. If we assume spectral
parameters characteristic of short GRBs, we still consistently find high
$\Gamma$ and low $n$.

Even though our results are not definitive, the strength of such an approach
lies in constraining values of the launching radius through GW observations and
address EM observations.

Further GW observations with better coverage from GBM will settle if merging BH
binaries indeed emit $\gamma$-rays.  It is possible however, that due to
observer angle effects, the GRB-GW association will only be settled once a
sizeable sample of GW and gamma-ray observations has accumulated. Indeed, GW
signals from compact mergers are not strongly dependent on the orientation of
the binary  while prompt gamma-rays are essentially not expected if we are not
inside the jet opening angle.  On the other hand, edge-on systems have on
average $1/\sqrt{8}$ the signal of the face on cases. This results in an
increased likelihood that the systems detected by Advanced LIGO are face-on
than edge-on.  By measuring the jet opening angle for a GRB, we can constrain
the available parameter space for the inclination, measured by the Advanced
LIGO. Furthermore, detailed multiwavelength afterglow modeling
\citep[e.g.][]{Zhang+15offaxis} can also constrain the viewing angle. 

Once GW observations become routine, and their EM counterparts will be readily
available, we will be able to address the question of the association of short
GRBs with BH mergers on a more solid footing. As an example of investigating
GW~150914-GBM as a member of the short GRB population, \citet{Li+16gw} argued
that it is an outlier on the $E_{\rm pk}-L_{\rm iso}$ diagram for short GRBs.
This may indicate a different progenitor for the GBM event. However, due to a
small sample size, the correlation for previous short GRBs is not strong enough
for a definite conclusion.

With our current understanding of GRBs, variability timescale of
the GRB lightcurve can provide a limit on the size of the jet launch.  If the
GW-GRB association is confirmed, the variability times can be compared to the
marginally stable radii resulting from the GW detection.  Thus it will be
possible to rule out some classes of models more firmly with similar analysis
to the one presented here.

{\it Acknowledgements -} We thank Tyson Littenberg and Michael Briggs for
discussions.  This study was supported by Fermi grant NNM11AA01A.  P.M.
acknowledges support from NASA NNX13AH50G.

\bibliographystyle{apj}
%-------

%-------
%\bibliography{gwgrb,\bibpth grb,\bibpth magnetic}
\end{document}